\documentclass[runningheads]{llncs}
\usepackage{graphicx}
\usepackage{booktabs}
\usepackage{amsmath}
\usepackage[flushleft]{threeparttable}
\usepackage{hyperref}
\begin{document}
\title{Automatic Tumor Segmentation via False Positive Reduction Network for Whole-Body Multi-Modality PET/CT Images}
\titlerunning{Automatic Tumor Segmentation via False Positive Reduction Network}
%
\author{Yige Peng\inst{1} \and
Jinman Kim\inst{1} \and
Dagan Feng\inst{1,2} \and
Lei Bi\inst{1}}
%
\authorrunning{Y. Peng et al.}
%
\institute{School of Computer Science, University of Sydney, NSW, Australia\and
Med-X Research Institute, Shanghai Jiao Tong University, Shanghai, China\\}
\maketitle              
\begin{abstract}
Multi-modality Fluorodeoxyglucose (FDG) positron emission tomography / computed tomography (PET/CT) has been routinely used in the assessment of common cancers, such as lung cancer, lymphoma, and melanoma. This is mainly attributed to the fact that PET/CT combines the high sensitivity for tumor detection of PET and anatomical information from CT. In PET/CT image assessment, automatic tumor segmentation is an important step, and in recent years, deep learning based methods have become the state-of-the-art. Unfortunately, existing methods tend to over-segment the tumor regions and include regions such as the normal high uptake organs, inflammation, and other infections. In this study, we introduce a false positive reduction network to overcome this limitation. We firstly introduced a self-supervised pre-trained global segmentation module to coarsely delineate the candidate tumor regions using a self-supervised pre-trained encoder. The candidate tumor regions were then refined by removing false positives via a local refinement module. Our experiments with the MICCAI 2022 Automated Lesion Segmentation in Whole-Body FDG-PET/CT (AutoPET) challenge dataset showed that our method achieved a dice score of 0.9324 with the preliminary testing data and was ranked 1st place in dice on the leaderboard. Our method was also ranked in the top 7 methods on the final testing data, the final ranking will be announced during the 2022 MICCAI AutoPET workshop. Our code is available at: \url{https://github.com/YigePeng/AutoPET_False_Positive_Reduction}.

\end{abstract}
\section{Introduction}
Multi-modality Fluorodeoxyglucose (FDG) positron emission tomography / computed tomography (PET/CT) is regarded as the imaging modality of choice for the diagnosis, staging, and treatment response observation of many cancers, such as lung cancer, lymphoma, and melanoma \cite{hatt_characterization_2017}. This is attributed to the fact that PET/CT combines the high sensitivity of PET in detecting regions of abnormal function and the specificity of CT in depicting the underlying anatomy of where the abnormal functions are occurring \cite{peng2021predicting}. Automatic tumor segmentation is an important prerequisite for quantitative PET/CT image analysis which enables tumor characterization, oncologic staging, and image-based therapy response assessment. Deep learning based technologies have made great progress in automatic medical image analysis \cite{peng2020multi}, and are regarded as the state-of-the-art in PET/CT tumor segmentation \cite{fu2021multimodal}. However, it remains challenging to get accurate tumor segmentation results. This is mainly attributed to the fact that tumors across different patients can have large variations in spatial locations, texture, shape, and appearance information. In addition, multiple tumors may present next to the normal high uptake regions, such that existing automatic segmentation methods tend to over-segment the tumor regions and include regions such as the normal high uptake organs, inflammation, and other infections as part of tumor regions. Therefore, a tumor segmentation model that can precisely delineate the tumor area and avoids false positive annotations is highly desirable for PET/CT image analysis.

The MICCAI 2022 Automated Lesion Segmentation in Whole-Body FDG-PET/CT (AutoPET) challenge provides a large training dataset to promote research on machine learning-based automatic tumor lesion segmentation  \cite{gatidis_automated_2022}. There are two specific requirements for the tumor lesion segmentation task: (1) accurate and fast lesion segmentation; (2) avoidance of false positive segmentations (e.g., brain, bladder, etc.). 

To address the tasks for the AutoPET challenge, we propose a false positive reduction network that accurately delineates the tumor regions in whole-body PET/CT images. We first introduce a self-supervised pre-trained global segmentation module to coarsely delineate the candidate tumor regions, then the candidate tumor regions are then refined by removing false positives via a local refinement module.

\section{Method}
\subsection{Materials}
AutoPET consists of a training dataset of 1,014 PET/CT scans derived from 900 patients acquired at the University Hospital Tübingen, Germany \cite{gatidis_whole-body_2022}. All images are in NifTI format. There are 513 scans without lesions, and 188, 168, and 145 scans are histologically proven with malignant melanoma, lung cancer, and lymphoma, respectively. In addition, all patients have clinical reports including cancer diagnosis, sex, and age. A separate test dataset was not released to the public and was only used for the evaluation. The test dataset has a preliminary test set of 5 studies for self-evaluation and a final test set of 200 studies for final ranking. The preliminary test set was part of the final test set where 100 studies were from the same hospital as the training database (University Hospital Tübingen) and the other 100 scans were acquired from the University Hospital of the LMU in Munich with a similar acquisition protocol. The tumor regions for all the training and testing datasets were annotated by two radiologists with more than 5 years of experience in Hybrid Imaging and experience in machine learning research. In this study, we did not use any external dataset to build and train our model.

\subsection{Data Pre-processing}
Multiple pre-processing steps were applied. Firstly, to compress the usage of GPU memory, all the PET/CT image volumes were cropped into a patch size of 224×224 in the axial plane. Then the images were set to the SUV range of [0, 14.25] for PET and HU range of [-800, 400] for CT, and further mapped to [0, 1] via min-max normalization. Finally, for PET images, the input slices were normalized with the mean and standard deviation values of the entire training dataset, such that to adjust all the regions of interest (ROIs) to a notionally common scale based on the metabolism intensity of tumor regions. For the CT images, the input slices were normalized with the mean and standard deviation values of the individual patient.

\begin{figure}
\centering
\includegraphics[width=\textwidth]{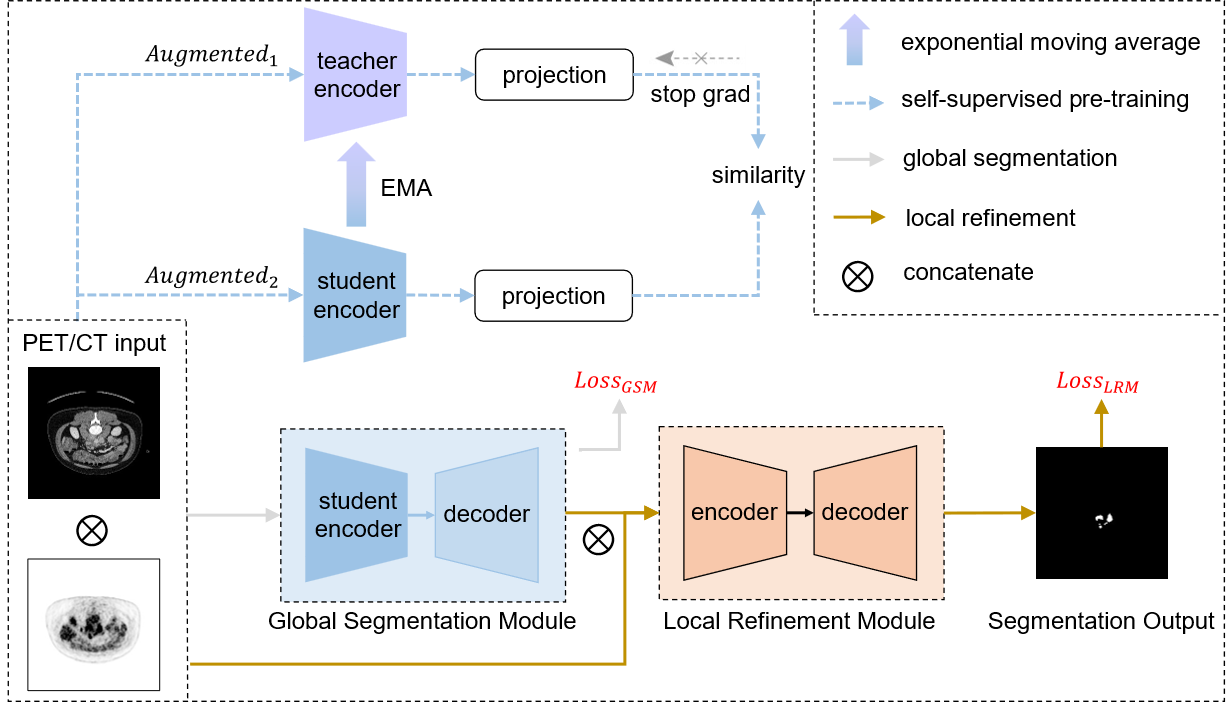}
\caption{The overview of our false positive reduction network. The arrows in different colors indicate different steps which are taken sequentially. The self-supervised pre-training improves the representation ability of tumor regions in whole-body PET/CT images; this is followed by the global segmentation which uses the pre-trained ResNet50 encoder to coarsely delineate the candidate regions. Afterward, the local refinement module removes the false positive regions using the output of the global segmentation module that is concatenated with the paired PET/CT images as input.} \label{fig1}
\end{figure}

\subsection{False Reduction Network}
Our false positive reduction network consists of two main modules, as shown in Fig. 1: a global segmentation module and a local refinement module. 

Specifically, a ResNet50 \cite{he_deep_2016} encoder is firstly pre-trained on the training data (tumor presented) via contrastive learning \cite{caron2021emerging} using concatenated 3-channel PET/CT images as the input. Two channels of the input images are set to be PET while the rest of 1 channel is assigned to CT. Then, a U-Net based decoder combined in together with the pre-trained encoder forms the global segmentation module, where a combined loss $\operatorname{Loss}_{GSM}$ is used for training. $\operatorname{Loss}_{GSM}$ consists of a dice loss \cite{milletari_v-net_2016} and a cross-entropy loss \cite{bishop_pattern_2006}, which is defined as:

\begin{equation}
\operatorname{Loss}_{GSM}=-\frac{2 \sum_i^N p_i g_i}{\sum_i^N p_i^2+\sum_i^N g_i^2}-\frac{1}{N} \sum_{i=1}^N\left[g_i \log p_i+\left(1-g_i\right) \log \left(1-p_i\right)\right]
\end{equation}

where $p_i \in \left[ 0,1 \right]$ is pixels of the predicted tumor probability map, $g_i \in \left[ 0,1 \right]$  is the pixels of the ground truth tumor mask (label), and the sums run over all available N pixels of the segmentation.

The global segmentation module can coarsely annotate the tumor lesion regions with a probability map at a threshold of 0.5. Both the probability map and the binary segmentation prediction (thresholded results) are further concatenated with the corresponding input PET/CT images, which are fed into the local refinement module. 

The backbone of the local refinement module is a 2D U-Net \cite{ronneberger_u-net_2015} with 5-channel imaging data as input (i.e., paired PET/CT images, global tumor probability map, and global binary segmentation prediction). A pixel-wise mean squared error (MSE) loss and a cross-entropy loss \cite{bishop_pattern_2006} are combined as $\operatorname{Loss}_{LRM}$ and used to compare the predicted segmentation outputs with the tumor ground-truth mask such that the false positive regions can be removed from the global segmentation with:

\begin{equation}
\operatorname{Loss}_{LRM}=\frac{1}{N} \sum_{i=1}^N\left\{\left(g_i-p_i\right)^2-\left[g_i \log p_i+\left(1-g_i\right) \log \left(1-p_i\right)\right]\right\}
\end{equation}

\subsection{Implementation Details}
Our method was implemented with PyTorch \cite{paszke_automatic_2017} using one NVIDIA GeForce GTX 2080Ti GPU. Our model was initialized using the approach presented by He et al \cite{he_delving_2015}, and an adaptive-moment estimation with decoupled weight decay (AdamW) \cite{loshchilov_decoupled_2018} was used for network optimization. During the training phase, the batch size was set to 8 and the learning rate was set to 0.0001 using a cosine annealing schedule. Data augmentation techniques were in real-time to avoid overfitting. The used data augmentation techniques are random rotation (90°, 180°, or 270°) in the axial axis and randomly flipping in one of the two axes (sagittal and coronal). 

Furthermore, we only used PET/CT slices with tumors for the global segmentation module, while an equal number of PET/CT slices without tumors were sampled and added to the training of the local refinement module. All the training was terminated when no further change is in the total loss. In our method, the total loss was generally stable after 160 epochs. Our results on the testing set were obtained using an ensemble of four models trained on different splits of the training set. Three models were built using 3-fold cross-validation, the other one was built from the 3D full resolution nnU-Net \cite{isensee_nnu-net_2021}. For the segmentation output, the testing prediction was a weighted sum between the nnU-Net output and the direct average result of the 3 cross-validation models, where the weights are empirically set to 0.35 and 0.65 respectively.

\section{Results}
Three different metrics are used for tumor segmentation evaluation, they are: (1) foreground Dice score of segmented tumors; (2) volume of false positives that do not overlap with positives (=false positive volume, FPV); and (3) volume of positive connected components in the ground truth that do not overlap with the estimated segmentation (=false negative volume, FNV).

For the test set evaluation (inaccessible to the participants), all three metrics are considered for non-healthy patients (tumors presented), whereas only FPV is considered for healthy cases (no tumors presented). The ﬁnal leader board position is based on the ranking across all three metrics with the weights of (0.5, 0.25, 0.25).

\begin{table}
\begin{threeparttable}
\centering
\caption{The segmentation results for the preliminary testing dataset}\label{tab1}
\begin{tabular*}{0.9\textwidth}{@{\extracolsep{\fill}}lccc@{}}
\toprule
Methods & Dice $\uparrow$ & FPV $\downarrow$ & FNC $\downarrow$\\
\midrule
Global Segmentation Module & 0.9228	& 0.9555 & 1.7865\\
Local Refinement Module & 0.9271	& 0.8360 & 1.7865\\
Ensemble with nnU-Net & \textbf{0.9324} & \textbf{0.7763} & \textbf{1.5676} \\
\bottomrule
\end{tabular*}
\begin{tablenotes}

\item The bold numbers represent the best results.
\item The arrows next to the evaluation metrics indicate how to assess the performance, $\uparrow$ means that the larger the number is, the better the performance is, and vice versa.
\end{tablenotes}
\end{threeparttable}
\end{table}

Our results in the preliminary testing dataset are presented in Table 1. With the global segmentation module only, our model achieved a Dice score of 0.9228, an FPV of 0.9555, and an FNV of 1.7865. Then, the inclusion of the local refinement module effectively removed the false positive regions, such that improved the FPV by a large margin of 12.5\% while the FNV was consistent with a minor improvement in dice score. The false positive reduction network ensembled with nnU-Net obtained overall better performance with a dice score, FPV, and FNV of 0.9324, 0.7763, and 1.5676 respectively, which was ranked 1st place in dice and 2nd place across all metrics on the leaderboard.

\section{Conclusion}
We introduced a false positive reduction network for the MICCAI 2022 AutoPET challenge. Our proposed method consists of two modules, namely the global segmentation module and the local refinement module. Within the global segmentation module, a self-supervised pre-trained encoder was designed to coarsely delineate the candidate tumor regions with a U-Net decoder, then the candidate tumor regions were refined by removing false positive segmentations via the local refinement modules. Our method achieved the highest dice score of 0.9324 with the preliminary test data on the leaderboard.

%
%
%
\bibliographystyle{splncs04}
\bibliography{mybibliography}

\end{document}